\documentclass[pra,twocolumn,showkeys,showpacs,superscriptaddress,floatfix,longbibliography]{revtex4-1}

\usepackage[T1]{fontenc}
\usepackage[latin9]{inputenc}
\usepackage{epsfig,psfrag}
\usepackage{latexsym}
\usepackage{graphicx}
\usepackage{dcolumn}
\usepackage{amssymb}
\usepackage{amsmath}
\usepackage{bm}
\usepackage{mathrsfs}
\usepackage{bigints}
\usepackage{upgreek}
\usepackage{color}
\usepackage{longtable}
\usepackage{xspace}
\usepackage[squaren]{SIunits}
\usepackage{epstopdf}
\usepackage[normalem]{ulem}
\usepackage{hyperref}

\definecolor{mygreen}{RGB}{20,148,20}

\usepackage{filemod}

\newcommand{\ie}{i.e.\@\xspace}

\newcommand{\eq}[1]{Eq.~\eqref{eq:#1}}
\newcommand{\eqs}[2]{Eqs.~\eqref{eq:#1} and~\eqref{eq:#2}}

\newcommand{\fig}[1]{Fig.~\ref{fig:#1}}

\newcommand{\Pel}{P_{\textrm{el}}}
\newcommand{\Psca}{P_{\textrm{sca}}}

\newcommand{\vbar}{\bar{v}}

\begin{document}

\title{Resonant forward-scattered field in the high-saturation regime: Elastic and inelastic contributions}


\author{C. C. Kwong}
\email{changchikwong@ntu.edu.sg}
\affiliation{Nanyang Quantum Hub, School of Physical and Mathematical Sciences, Nanyang Technological University, 21 Nanyang Link, Singapore 637371, Singapore.}
\affiliation{MajuLab, International Joint Research Unit UMI 3654, CNRS, Universit\'e C\^ote d'Azur, Sorbonne Universit\'e, National University of Singapore, Nanyang Technological University, Singapore.}
\author{T. Wellens}
\affiliation{Physikalisches Institut, Albert-Ludwigs-Universit{\"a}t, Hermann-Herder-Strasse 3, D-79104 Freiburg, Germany.}
\author{K. Pandey}
\affiliation{Department of Physics, Indian Institute of Technology Guwahati, Guwahati, Assam 781039, India.}
\author{D. Wilkowski}
\affiliation{Nanyang Quantum Hub, School of Physical and Mathematical Sciences, Nanyang Technological University, 21 Nanyang Link, Singapore 637371, Singapore.}
\affiliation{MajuLab, International Joint Research Unit UMI 3654, CNRS, Universit\'e C\^ote d'Azur, Sorbonne Universit\'e, National University of Singapore, Nanyang Technological University, Singapore.}
\affiliation{Centre for Quantum Technologies, National University of Singapore, 117543 Singapore, Singapore.}


\begin{abstract}
We measure the resonant forward scattering of light by a highly saturated atomic medium 
through the flashes
emitted immediately after an abrupt extinction of the probe beam. The experiment is done in a dilute regime where the phenomena are well captured using the independent scattering approximation. Comparing our measurements to a model based on Maxwell-Bloch equations, our experimental results are consistent with contributions from only the elastic component, whereas the attenuation of the coherent transmission power is linked to the elastic and inelastic scatterings. In the large saturation regime and at the vicinity of the atomic resonance, we derive an asymptotic expression relating the elastic scattering power to the forward-scattered power.
\end{abstract}

\pacs{}

\keywords{}

\maketitle
\section{Introduction}

When a quasiresonant laser beam is shined on an ensemble of atomic 
emitters,
the light undergoes scattering by the randomly positioned atoms. This process depletes photons from the incoming beam, resulting in an attenuation of the power in the coherent transmission. At low intensity, the scattering events are elastic with well-defined phase such that the process remains coherent. Light scattering and transport become more complex at high incident intensity~\cite{binninger2018}, when the atomic transition becomes saturated. Inelastic scattering contributions lead to the well-known Mollow triplet~\cite{mollow1969,cohen2004}. It also leads to an additional incoherent background to the coherent transport of light, which causes a reduction in the contrast of coherent backscattering~\cite{wilkowski2004, chaneliere2004,PhysRevA.70.023817,balik2005}. Due to  the saturation effect, the transmission of light  becomes nonlinear. It leads to the phenomenon of self-induced transparency~\cite{mccall1969}, which provides one method where the optical precursors~
\footnote{For a pulse travelling through a dispersive medium, the optical precursor is the front of the pulse which travels at the vacuum speed of light~\cite{sommerfeld1914, brillouin1914}. It has been observed in solid state~\cite{aaviksoo1991} and cold atomic systems~\cite{jeong2006, wei2009}. } can be  separately measured from the main pulse~\cite{macke2010,marskar2012}. 

In the steady-state regime, energy conservation states that the optical power attenuated in the coherent transmission is converted to the total scattered power, which consists
of both 
elastic and inelastic contributions. In addition, if one considers a beam falling on a slab, the superposition principle, in the far field along the forward direction, leads to $E_t=E_0+E_s$.
Here, the transmitted field $E_t$ is interpreted as a coherent superposition between the incident field $E_0$ and the forward-scattered field $E_s$,
the latter being a coherent field built up on elastic scattering only. This reasoning leads to the remarkable result that the
forward-scattered
field is governed by elastic events, whereas
the flux of photons scattered into other directions is
linked to the total scattering events, including both elastic and inelastic processes. 

A direct measurement of the forward-scattered light in the steady-state is hindered by its superposition with the incident field. Fortunately, in the transient regime, the measurement of the steady state $E_s$ becomes accessible. In particular, the flash effect~\cite{chalony2011,kwong2014} has been used to experimentally 
measure the forward-scattered field $E_s$ in the linear regime. The coherent emission of a flash is achieved by abruptly switching off  the incident probe beam, so that the atoms in the medium undergo free induction decay (FID)~\cite{hahn1950,brewer1972,toyoda1997,shim2002}. Related to these works are studies performed in the transient regime during probe ignition (some examples include optical precursor~\cite{jeong2006, wei2009, macke2010}, stimulated inelastic resonance fluorescence~\cite{makarov1979,eberly1980,macke1981,segard1981}, and initial flashes~\cite{chalony2011}). When the medium has a large optical depth, the FID appears as a flash of light during probe extinction, with a time-scale shorter than the natural lifetime of the transition~\cite{kwong2015}. Since the response time of the atoms is finite, $E_s$ remains continuous across any abrupt change to the probe beam. Thus, a detector placed in the exact forward direction will initially measure an optical power $P_s$ that is associated with the field $E_s$. The phase of $E_s$, relative to $E_0$, can be extracted by measuring the incident power $P_0$ and the steady-state transmitted power $P_t$~\cite{chalony2011}, or by abrupt phase variation of the incident field~\cite{kwong2015}.

In this article, we analyze the emission of flashes in the saturated regime at large optical depth. We 
perform
experimental measurements of the steady-state transmittance and peak flash power as functions of  probe beam detuning and saturation parameter. The peak flash power gives a direct measurement of the forward-scattered power. We compare the experimental results with a model based on the Maxwell-Bloch equation, showing that only the elastic component contributes to the forward-scattered field. The Maxwell-Bloch equation is commonly used to study the propagation of light through atomic media (see a recent work in Ref.~\cite{jennewein2018}). We then discuss at large saturation parameter, how the forward-scattered power can allow us to determine the elastic scattering power. The paper is organized as follows. In Sec.~\ref{sec:model}, we derive the Maxwell-Bloch equations using a two-level model, focusing on the calculation of  experimentally measured quantities such as the transmitted power and the forward-scattered power. The experimental results are presented  and compared to the theoretical predictions in Sec.~\ref{sec:expt}.  In Sec.~\ref{sec:discussion}, we discuss the link between the forward-scattered power and elastic scattering power, in the large saturation regime.

\section{Theoretical model}\label{sec:model}

In this section, we present our model to describe the coherent transmission of light in the saturated regime. The coherent forward-propagating field inside a two-level atomic cloud can be written as a sum of two fields,
\begin{equation}
E_t({\bf r},t)=E_0({\bf r},t)+E_s({\bf r},t),\label{eq:Et}
\end{equation}
where ${\bf r}=(x,y,z)$, with $z$ directed along the propagation direction.
The first term on the right-hand side corresponds to the incident laser field,
\begin{equation}
E_0({\bf r},t)=E_0(x,y)e^{i k z}\Theta(t_{\rm sw}-t),\label{eq:E0}
\end{equation}
which is suddenly switched off at time $t_{\rm sw}$.  The wave 
number
of the probe beam is denoted by $k$. For simplicity, we neglect the finite propagation time of the light through the atomic medium. $\Theta$ is the unit step function. 
The amplitude $E_0(x,y)=E_0 \exp[-(x^2+y^2)/w_0^2]$ exhibits a Gaussian profile with a beam waist of $w_0$. We assume that the Rayleigh length is much larger than the cloud size, so that the waist can be taken to be constant inside the cloud.
The maximum amplitude $E_0$ is related to a saturation parameter at resonance $s_0=2|\Omega_0|^2/\Gamma^2$, where $\Omega_0 = -E_0d/\hbar$ and  $d^2 = 3\pi\epsilon_0\hbar\Gamma/k^3$. $\Gamma$ is the transition linewidth, $d$ is the reduced electric dipole moment, $\hbar$ is the reduced 
Planck
constant, and $\epsilon_0$ is the vacuum permittivity.

The last term in \eq{Et}, $E_s({\bf r},t)$, denotes the field which is coherently scattered by the atoms in the forward direction. The atomic density (see Sec.~\ref{sec:setup}) is low enough to put us in the  dilute regime of light scattering. We make the assumption that each atom scatters light independently.  Upon neglecting the propagation time, we obtain
\begin{equation}
E_s({\bf r},t)=-\frac{3 i \pi \hbar\Gamma}{dk^2 }\int_{-\infty}^z {\rm d}z'~\rho({\bf r}') e^{i k (z-z')} \sigma_-({\bf r}',t),\label{eq:scattered}
\end{equation}
where ${\bf r}'=(x,y,z')$ and 
\begin{equation}
\rho({\bf r})=\rho \exp\left[-\frac{x^2}{2R_\parallel^2}-\frac{y^2+z^2}{2 R_\perp^2}\right]
\end{equation}
 denotes the density profile of the ellipsoidal cloud. $R_\parallel$ ($R_\perp$) is the axial (equatorial) radius of the cloud. 
$\sigma_-({\bf r}',t)$ refers to the first component of the atomic Bloch vector (see below), which gives rise to coherent scattering of light~\cite{cohen2004}.

\subsection{Bloch equations}

With strontium atoms laser-cooled on the narrow intercombination line (see Sec.~\ref{sec:setup}), the resulting cold atomic cloud still experiences residual Doppler broadening. To take this into account, the component $\sigma_-({\bf r},t)$ of the atomic Bloch vector is calculated with an average over the atomic velocity distribution $g(v)$,
\begin{equation}
\sigma_-({\bf r},t)=\int_{-\infty}^\infty {\rm d}v~g(v)\sigma_-^{(v)}({\bf r},t).
\end{equation}
The velocity distribution is a Gaussian distribution with a standard deviation $\vbar$, 
\begin{equation}
g(v) = \frac{1}{\sqrt{2\pi}\vbar}\exp\left(-\frac{v^2}{2\vbar^2}\right).
\end{equation}

For a given velocity $v$, the atomic Bloch vector fulfills the following optical Bloch equations in the rotating wave approximation:
\begin{widetext}
\begin{eqnarray}
\frac{{\rm d}}{{\rm d}t} \sigma_-^{(v)}({\bf r},t) & = & \left(i(\delta-k v)-\frac{\Gamma}{2}\right)\sigma_-^{(v)}({\bf r},t)-i \frac{\Omega_t({\bf r},t)}{2}\sigma_z^{(v)}({\bf r},t), \nonumber\\
\frac{{\rm d}}{{\rm d}t} \sigma_+^{(v)}({\bf r},t) & = & \left(-i(\delta-k v)-\frac{\Gamma}{2}\right)\sigma_+^{(v)}({\bf r},t)+i \frac{\Omega_t^*({\bf r},t)}{2}\sigma_z^{(v)}({\bf r},t), \nonumber\\
\frac{{\rm d}}{{\rm d}t} \sigma_z^{(v)}({\bf r},t) & = & - i \Omega_t^*({\bf r},t) \sigma_-^{(v)}({\bf r},t)+i \Omega_t({\bf r},t) \sigma_+^{(v)}({\bf r},t)
-\Gamma \left[\sigma_z^{(v)}({\bf r},t)+1\right].\label{eq:bloch}
\end{eqnarray}
\end{widetext}
The local Rabi frequency is denoted by $\Omega_t({\bf r},t)=-E_t({\bf r},t) d/\hbar$, where $E_t({\bf r},t)$ is given by Eqs.~(\ref{eq:Et})--(\ref{eq:scattered}).
In Eq.~(\ref{eq:bloch}), we disregard the change of atomic velocities caused by collisions between the atoms (not relevant in the regime of  small density and low temperatures realized in our experiment) and recoils due to the scattering of photons (since we are concerned with the forward-scattered fields).

The total  power of the coherently transmitted light, integrated over the laser beam transverse profile, reads
\begin{equation}
P_t(t)=\frac{c_0\epsilon_0}{2}\int_{-\infty}^\infty{\rm d}x\int_{-\infty}^\infty {\rm d}y~|E_t(x,y,\infty,t)|^2,\label{eq:Jt}
\end{equation}
where $c_0$ is the speed of light in vacuum. We derive optical powers instead of intensities, since the incident light has a transverse Gaussian profile, and the photodetector effectively integrates over the intensities in this transverse direction. Moreover, the transverse beam profile should be properly taken into account because of the nonlinear response of the atomic medium. We also assume that the light rays propagate parallel to the optical axis, $z$, disregarding the linear and nonlinear focusing or defocusing effect on the beam due to transverse gradients of the medium refractive index. We normalize the transmitted power to the total incident power,
\begin{equation}
P_0=\frac{c_0\epsilon_0}{2}\int_{-\infty}^\infty{\rm d}x\int_{-\infty}^\infty {\rm d}y~|E_0(x,y)|^2=\frac{\pi c_0\epsilon_0 w_0^2 |E_0|^2}{4},
\end{equation}
and we introduce the on-resonant optical depth up to the point $z$ in the cloud,
\begin{equation}
\zeta({\bf r})=\int_{-\infty}^z\frac{{\rm d}z'}{\ell_0({\bf r}')}=\frac{6\pi}{k^2}\int_{-\infty}^z {\rm d}z'~\rho({\bf r}').\label{eq:Zeta}
\end{equation}
We stress that the above optical depth is defined in the absence of Doppler broadening. Furthermore, $\ell_0({\bf r}')=1/[\sigma_0\rho({\bf r}')]$,  denotes the mean free path in a dilute medium of point scatterers. $\sigma_0=6\pi/k^2$ is the on-resonance scattering cross section of light. The corresponding resonant optical depth is
\begin{equation}
b_0=\lim_{z\to\infty} \zeta({\bf r}).
\end{equation}
It is considered to be independent of the transverse coordinates, since the beam waist is smaller than the smallest diameter of the atomic cloud ellipsoid (see Sec.~\ref{sec:setup}). This allows us to approximate the geometry of the medium as a slab. Our problem becomes rotationally invariant around the optical axis ($z$ axis). The off-center parts of the beam are taken into account by a transverse-dependent saturation parameter,
\begin{equation}
s(x,y)=s_0 \exp\left[-2\frac{x^2+y^2}{w_0^2}\right]. \label{eq:s0andR}
\end{equation}
Moreover, we introduce dimensionless fields with constant propagation phase as
follows:
\begin{eqnarray}
{\mathcal E}_t(\zeta,s,t)& = & \frac{E_t({\bf r},t)}{E_0(x,y)} e^{-ikz},\\
{\mathcal E}_s(\zeta,s,t)& = & \frac{E_s({\bf r},t)}{E_0(x,y)} e^{-ikz}.
\end{eqnarray}
Similarly, the atomic Bloch vector is rescaled according to
\begin{eqnarray}
\tilde{\sigma}_\mp^{(v)}(\zeta,s,t) & = & -\frac{\hbar\Gamma \sigma_\mp^{(v)}({\bf r},t)}{d E_0(x,y)}  e^{\mp i k z},\\
\tilde{\sigma}_z^{(v)}(\zeta,s,t) & = & \sigma_z^{(v)}({\bf r},t)
\end{eqnarray}
We use the same rescaling for the velocity-averaged quantities, and Eqs.~(\ref{eq:Et})-(\ref{eq:scattered}) become
\begin{eqnarray}
{\mathcal E}_t(\zeta,s,t) & = & \Theta(t_{\rm sw}-t)+{\mathcal E}_s(\zeta,s,t),\label{eq:Et_mod}\\
{\mathcal E}_s(\zeta,s,t)& = & \frac{i}{2}\int_0^{\zeta}{\rm d}\zeta'~\tilde{\sigma}_-(\zeta',s,t).
\label{eq:scattered_mod}
\end{eqnarray}
The correspondingly modified Bloch equations are 
\begin{widetext}
\begin{eqnarray}
\frac{1}{\Gamma}\frac{{\rm d}}{{\rm d}t} \tilde{\sigma}_-^{(v)}(\zeta,s,t) & = & \left(\frac{i}{\Gamma}(\delta-k v)-\frac{1}{2}\right)\tilde{\sigma}_-^{(v)}(\zeta,s,t)-i \frac{{\mathcal E}_t(\zeta,s,t)}{2}\tilde{\sigma}_z^{(v)}(\zeta,s,t) \nonumber\\
\frac{1}{\Gamma}\frac{{\rm d}}{{\rm d}t} \tilde{\sigma}_+^{(v)}(\zeta,s,t) & = & \left(-\frac{i}{\Gamma}(\delta-k v)-\frac{1}{2}\right)\tilde{\sigma}_+^{(v)}(\zeta,s,t)+i \frac{{\mathcal E}_t^*(\zeta,s,t)}{2}\tilde{\sigma}_z^{(v)}(\zeta,s,t) \nonumber\\
\frac{1}{\Gamma}\frac{{\rm d}}{{\rm d}t} \sigma_z^{(v)}(z,s,t) & = & -1 - \frac{i s}{2}  {\mathcal E}_t^*(\zeta,s,t) \tilde{\sigma}_-^{(v)}(\zeta,s,t)+\frac{i s}{2} {\mathcal E}_t(\zeta,s,t) \tilde{\sigma}_+^{(v)}(\zeta,s,t)
-\tilde{\sigma}_z^{(v)}(\zeta,s,t)\label{eq:modbloch}
\end{eqnarray}
\end{widetext}

The total coherently transmitted power is obtained by rewriting Eq.~(\ref{eq:Jt}) in the new variables,
\begin{equation}
\frac{P_t(t)}{P_0}=\int_0^{s_0} \frac{{\rm d}s}{s_0}~\left|{\mathcal E}_t\left(b_0,s,t\right)\right|^2.\label{eq:Jt_mod}
\end{equation}

If $t_{\rm sw} \gg 1/\Gamma$, a steady-state regime is achieved before the probe beam is switched off. This regime is obtained by setting the time derivatives in Eq.~(\ref{eq:modbloch}) to zero, thereby
expressing the Bloch vector as an analytical function of the field ${\mathcal E}_t$.
The resulting integral in \eq{scattered_mod} can be solved by iteration. Using this solution in Eq.~(\ref{eq:Jt_mod}) yields the steady-state transmitted power.

Just after switching off the probe beam, the transmitted field jumps from
${\mathcal E}_t=1+{\mathcal E}_s$ to ${\mathcal E}_t={\mathcal E}_s$ [see Eq.~(\ref{eq:Et_mod})]. The corresponding  peak power of the flash is therefore obtained as:
\begin{equation}
\frac{P_s}{P_0}=\int_0^{s_0} \frac{{\rm d}s}{s_0}~\left|{\mathcal E}_s\left(b_0,s\right)\right|^2,\label{eq:Js_mod}
\end{equation}
where ${\mathcal E}_s\left(b_0,s\right)$ is the steady-state value of forward scattering.

%

\section{Experimental results}\label{sec:expt}

\subsection{Experimental setup and parameters}\label{sec:setup}
\begin{figure}
\includegraphics[width=0.5\textwidth]{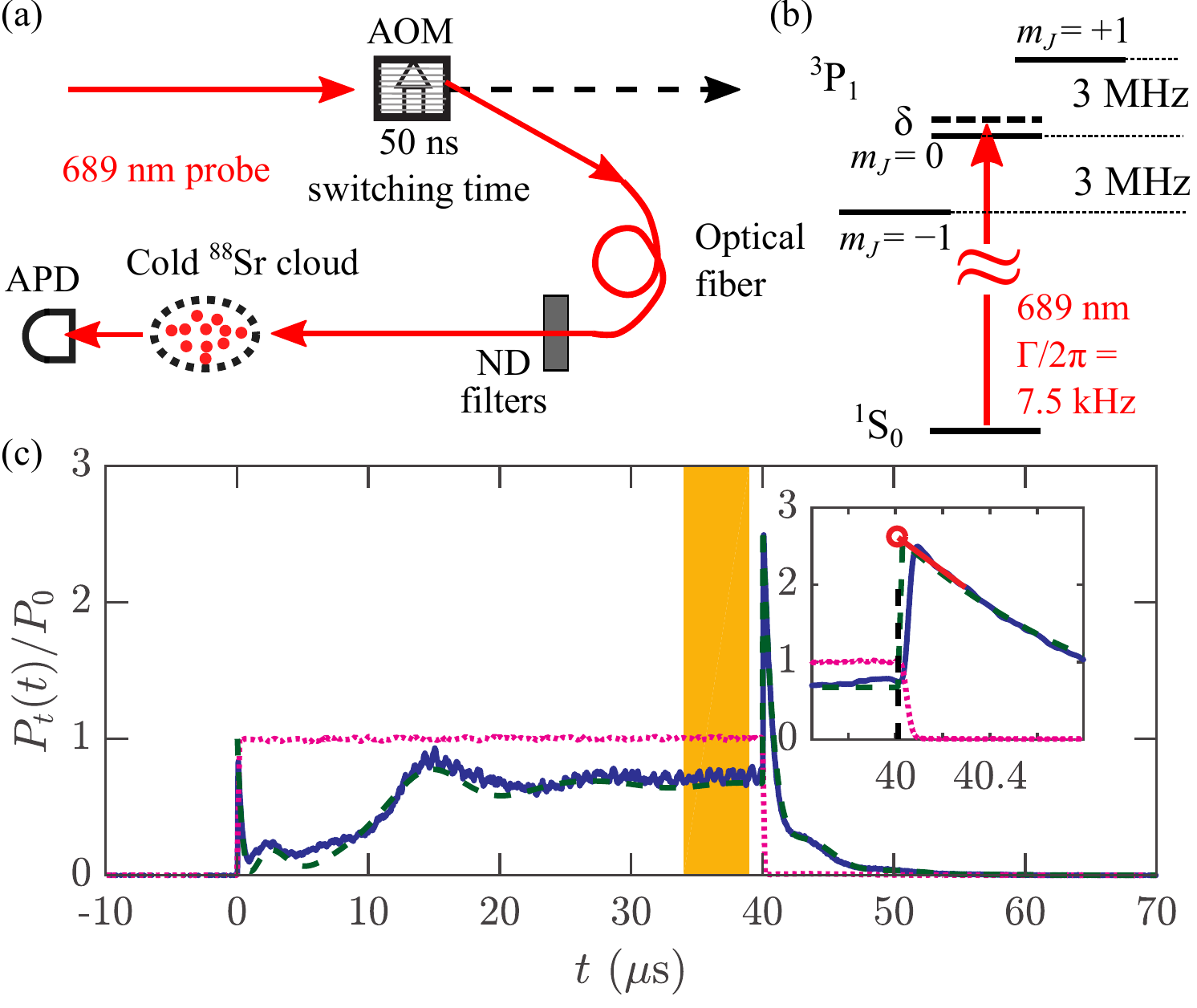}
\caption{(a) The experimental setup. (b) The $^{88}$Sr  intercombination line transition probed in the experiment. The detuning from the $m_J=0\rightarrow m_J=0$ transition is denoted by $\delta$.  (c) An example of a temporal signal at $s_0=297$ and $\delta = -60$~kHz ($\delta/\Gamma = -8$). The blue solid curve is the experimental data and the green dashed line is a theoretical calculation of the temporal behavior (see Sec.~\ref{sec:model}). The magenta dotted curve represents the incident probe beam. The orange shaded area indicates the temporal window used to compute the average transmitted power $P_t$ in the steady-state regime. 
The 
inset fshows the linear extrapolation (red line) in a zoom around the probe extinction. The extrapolated peak value of the flash emission
is indicated by the red open circle.}\label{fig:fig1}
\end{figure}

The experimental setup is sketched in \fig{fig1}(a). A $\lambda=689$~nm laser probes the {$^1$S$_0\rightarrow ^3$P$_1$} intercombination line transition of a cold $^{88}$Sr atomic ensemble, where the natural linewidth is $\Gamma/2\pi=7.5$~kHz. Cooling and trapping details of the $^{88}$Sr atoms are discussed in Ref.~\cite{yang2015}. In brief, atoms are laser cooled in a  magneto-optical trap to a final temperature of $T=3.3$~$\mu$K. The experiment is performed 10~ms after the atoms are released from the magneto-optical trap. The cloud  takes an ellipsoidal shape with an axial radius of $R_\parallel=240(10)~\mu$m along the vertical direction, and an equatorial radius of $R_\perp=380(30)~\mu$m. The cloud consists of 2.5(5)$\times10^8$ atoms, leading to a peak density of  $\rho=4.6\times10^{11}$~cm$^{-3}$. From the transmission measurement at low intensity, we find that the cloud has an optical depth of $b=19(3)$. This corresponds to a resonant optical depth at zero temperature of $b_0=115(10)$~\cite{kwong2014}. The two optical depths are related by $b = b_0g(k\vbar/\Gamma)$, where $g(x) = {\sqrt{\pi/8}\exp(1/8x^2)\rm{erfc}(1/\sqrt{8}x)/x}$. $\vbar$ is the thermal velocity of the atoms, defined by $\vbar = \sqrt{k_BT/m}$, with $k_B$ the Boltzmann factor and $m$ the atomic mass.

The atomic density in our experiments (see Sec. III A) is low enough such that the atoms mostly experience the far field of light scattered by other atoms. While superradiance and subradiance effects~\cite{dicke1954, gross1982} have been reported in dilute systems (see, e.g., \cite{roof2016,araujo2016,guerin2016}), we found that our experimental results can be well explained, keeping superradiance and subradiance effects out of the picture. Modifications of the light scattering properties due to dipole-dipole interactions~\cite{keaveney2012, balik2013, pellegrino2014, javanainen2014,jennewein2016, bromley2016, corman2017, kwong2019} are also not expected to play a role here.

We apply a 1.4~G magnetic field along the linear polarization direction of the probe beam. This lifts the degeneracy of the excited $^3$P$_1$ state, allowing us to probe a two-level system corresponding to the magnetically insensitive $m_J=0\rightarrow m_J=0$ transition of the intercombination line [see \fig{fig1}(b)]. The probe beam is focused to a waist of 150~$\mu$m, its Rayleigh length is $z_R=\pi w_0^2/\lambda\simeq 10~{\rm cm}$, thus satisfying the assumptions, $w_0\ll R_{\parallel},\,R_{\perp}$ and $z_R\gg R_{\perp}$, made in Sec.~\ref{sec:model}.  The power of the incident probe beam can be adjusted between 530~pW to 23~$\mu$W by applying different neutral density (ND) filters. For the intercombination line where the saturation intensity is $I_s=3$~$\mu$W cm$^{-2}$, the peak resonant saturation parameter $s_0$ of the Gaussian probe beam ranges from $0.5$ to $2.2\times10^4$. 

To limit the effect of radiation pressure force on the atoms, the probe beam is turned on for a duration of $t_{\rm sw}=40$~$\mu$s~$\approx1.9/\Gamma$. The probe duration might be too short to populate possible subradiant modes of the cloud~\cite{guerin2016}, but it is long enough to achieve steady-state transmission in the independent light scattering picture.  Using an acousto-optic modulator (AOM), the falling time of the probe beam during the switch-off is 50~ns. This is more than 400 times shorter than the $1/\Gamma=$21~$\mu$s lifetime of the intercombination line. We detect the transmitted light using a 10~MHz bandwidth avalanche photodiode (APD), effectively performing a transverse integral over the transmitted intensity. The measurement signal of the detector is proportional to the transmitted power $P_t(t)$. After the transmission measurement, the atoms are blown away by a strong 461~nm laser tuned on the dipole-allowed {$^1$S$_0\rightarrow ^1$P$_1$} transition. The probe beam is then turned on again, with the same duration, to record the incident power $P_0$. 

The experiment is repeated for different detuning and saturation parameter values  of the probe laser. \fig{fig1}(c) shows an example of the transmission signal $P_t(t)/P_0$  for $s_0=297$ and $\delta/\Gamma=-8$, where a superflash (\textit{i.e.} a flash with a normalized amplitude large than 1) is observed after extinction of the probe beam. The green dashed curve is the theoretical curve obtained by solving Eqs.~(\ref{eq:Et_mod}) and (\ref{eq:scattered_mod}) together with the Bloch equations (\ref{eq:modbloch}) using Runge-Kutta integration.

\subsection{Steady-state transmitted power}
The experimental value of the steady-state transmitted power $P_t$ is obtained by averaging the transmission signal in a temporal window of {$34~\mu\rm s < t < 39~\mu\rm s$} [the orange shaded area in \fig{fig1}(c)].

\begin{figure}
\includegraphics[width=0.5\textwidth]{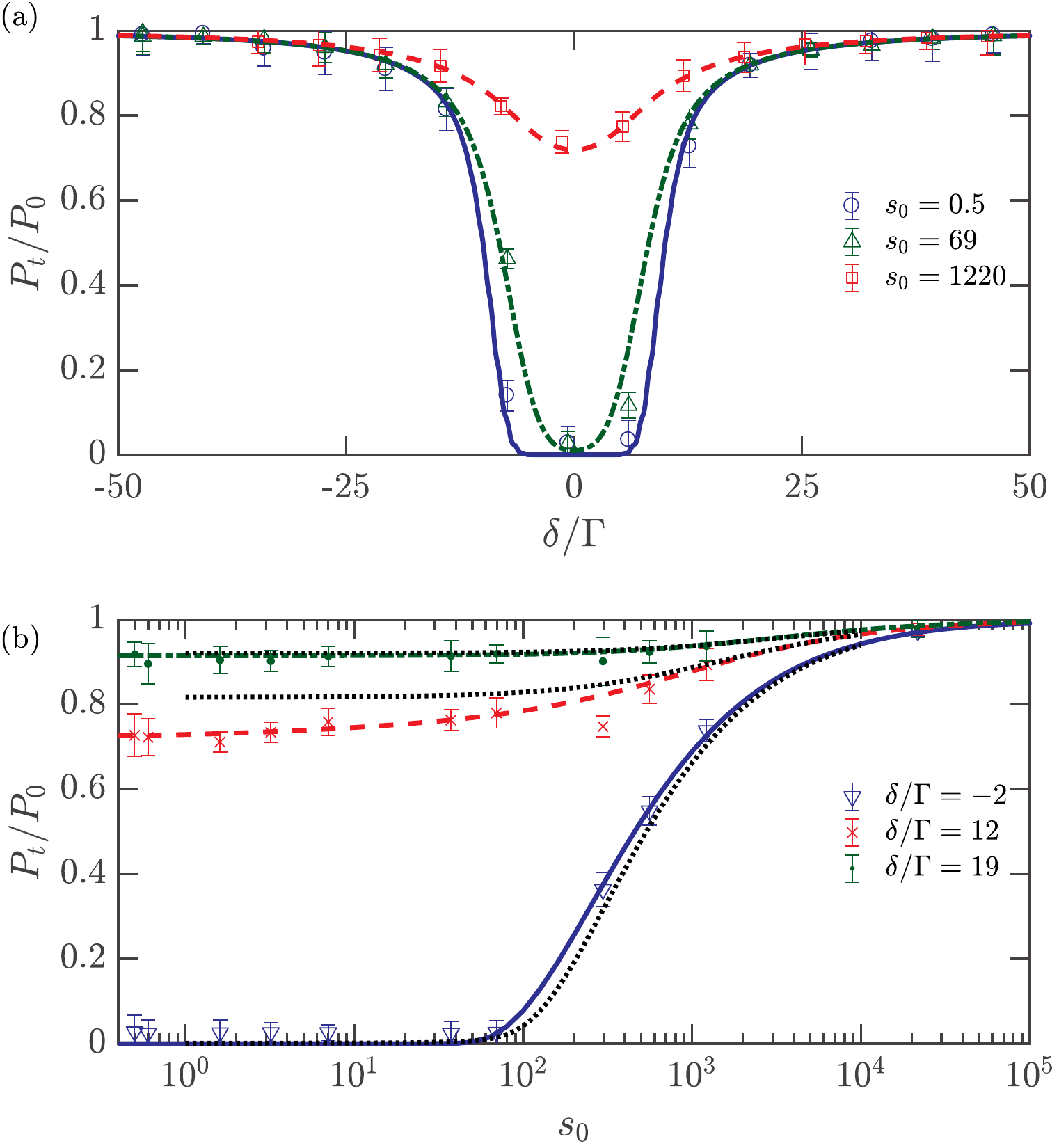}
\caption{ (a) The steady-state transmitted power as a function of the probe beam detuning, for various saturation parameters. The curves are the theoretical predictions; the blue solid curve is for $s_0=0.5$, the green dash-dotted curve is for $s_0=69$, and the red dashed curve is for $s_0=1220$. (b) The steady-state transmitted power as a function of the probe beam saturation parameter, plotted for three detuning values. The blue solid curve is for the theoretical prediction at $\delta/\Gamma = -2$, the red dashed curve is for the theoretical prediction at $\delta/\Gamma = 12$ and the green dash-dotted curve is for the theoretical prediction at $\delta/\Gamma=19$. The black dotted lines are the theoretical values of $P_t/P_0$, neglecting the Doppler broadening. In (a) and (b), the data points in the plots are the experimental data for different cases as indicated by the legends. The full range of the error bars represents two standard deviations.
	}\label{fig:fig2}
\end{figure}

In \fig{fig2}(a), we plot the measurements of $P_t$, at low ($s_0=0.5$), intermediate ($s_0=69$), and high  ($s_0=1220$) values  of the saturation parameter. In \fig{fig2}(b), the values of $P_t/P_0$ are plotted as a function of $s_0$, for a probe detuning that is near resonant ($\delta/\Gamma = -2$), at the low-intensity superflash regime ($\delta/\Gamma = 12$), and at the tail of the absorption window ($\delta/\Gamma = 19$). The curves are the predicted values using \eq{Jt_mod}, following the procedure  outlined just after that equation. The agreement between the theory and the experimental data is excellent.

We observe for the theoretical curves, that up to $s_0\sim 100$, the values of $P_t/P_0$ remain similar to the low saturation value. To explain this behavior, we consider the zero temperature case for a uniform beam of cross section area $A$. We find a simple transcendental equation that can be solved numerically for the saturation parameter of the transmitted probe beam, $s_t= P_t / (A I_s)$:
\begin{equation}
\left(\frac{4\delta^2}{\Gamma^2}+1\right)\ln\left(\frac{s_t}{s_0}\right)  = b_0\left[\frac{(s_0-s_t)}{b_0}-1\right],\label{eq:wtD}
\end{equation}
The above expression was derived from the modified Beer-Lambert law that accounts for the saturation of the transition~\cite{chaneliere2004, reinaudi2007, hueck2017}.  When $s_0-s_t \ll b_0$, \eq{wtD} simplifies to the linear case at low saturation. 
Therefore, saturation effects
start to take place for the majority of the atoms
when $s_0 \sim 100$. In this regime, power broadening of the atomic transition dominates over optical depth broadening of the absorption window. The same argumentation can be applied to our experiment,  with the presence of Doppler broadening [see \fig{fig2}(b)].  Above $s_0=100$, bleaching of the medium occurs, as indicated by the increasing values of $P_t$ as $s_0$ increases. Integrating the solution of \eq{wtD} over the transverse profile of the Gaussian beam, we find the three dotted lines in \fig{fig2}(b) for  $\delta/\Gamma=-2$, 12 and 19 respectively. We see that this zero-temperature case is a good approximation to the Doppler broadened case when either $s_t$ or $\delta$ dominates over the Doppler broadening $k\vbar/\Gamma$ (see Appendix~\ref{sec:zeroT} for more details). 

The statistical errors are represented by the error bars in \fig{fig2}. At low saturation, the experimentally measured transmission inside the absorption window is larger than expected, as observed in a previous study~\cite{kwong2014}. Since the atomic cloud has a finite size, a small fraction of light can be fully transmitted at the tails of the atomic cloud and captured by the photodetector. The transmittance threshold is found to be 2\%, confirmed by a simulation taking into account the actual sizes of the cloud and probe beam in the low saturation regime.
%

\subsection{Peak values of the flash}
\begin{figure}
\includegraphics[width=0.5\textwidth]{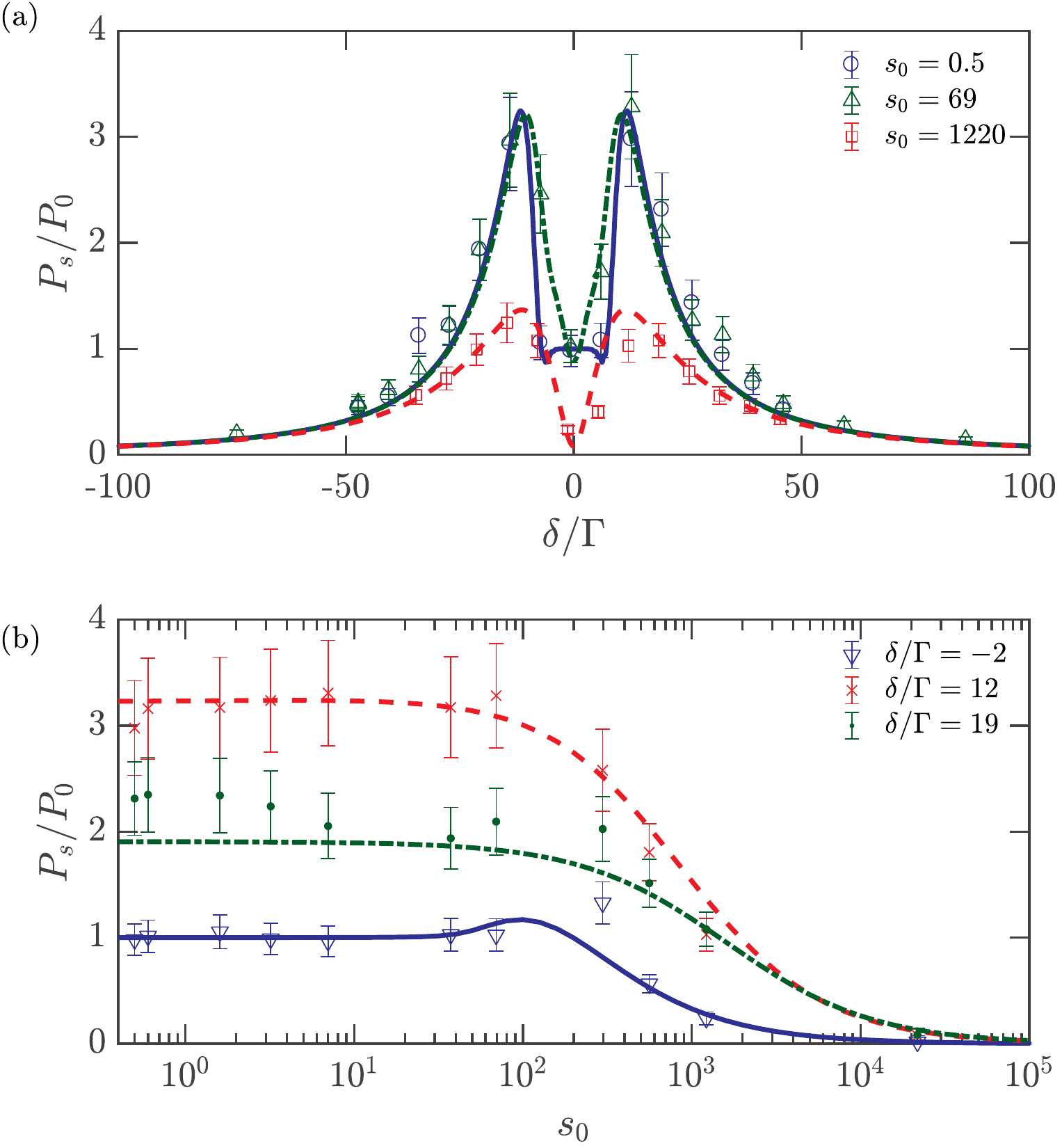}
\caption{(a) The  peak power of the flash as a function of probe beam detuning, for several saturation parameters of the probe beam. The curves are the theoretical predictions calculated using \eq{Js_mod}; the blue solid curve is for $s_0=0.5$, the green dash-dotted curve is for $s_0=69$ and the red dashed curve is for $s_0=1220$. (b) The peak value of the flash as a function of the probe beam saturation parameter, plotted for three detuning values. The blue solid curve is for the theoretical prediction at $\delta/\Gamma = -2$, the red dashed curve is for the theoretical prediction at $\delta/\Gamma = 12$ and the green dash-dotted curve is for the theoretical prediction at $\delta/\Gamma=19$. In (a) and in (b), the data points in the plots are the experimental data,
which, overall, agree very well with the theoretical predictions.	
	 }\label{fig:fig3}
\end{figure}

A flash is emitted in the forward direction, when the probe beam is turned off. The power of the forward transmission just after the extinction of the probe beam
is given by the steady-state forward-scattered power. In other words, $P_t(t_{\rm sw}) = P_s$. However, due to the finite response time of the detection scheme, the expected discontinuity in the transmission signal is smoothed out [see the inset of \fig{fig1}(c)].  A linear fit (red line) is applied to extrapolate the peak power of the flash $P_s/P_0$, at the time $t=t_{\rm sw}$ when the laser beam is switched off. Similar procedures were performed in Refs.~\cite{chalony2011,kwong2014}. The linear fit is done between $t=t_{\rm sw} + 50$~ns and $t=t_{\rm sw} + 200$~ns. In \fig{fig3}, we plot the peak power of the flash, $P_s/P_0$,
obtained from the linear extrapolation method.  In \fig{fig3}(a), $P_s/P_0$ is plotted against $\delta/\Gamma$ for three different saturation parameters $s_0=0.5$, 69, and 1220. In \fig{fig3}(b), $P_s/P_0$ is plotted against $s_0$ for three different probe detunings, $\delta/\Gamma=-2$, 12, and 19. 

Outside the absorption window, superflashes are emitted when the phase shift of the transmitted field becomes out of phase with the incident field. We measure a maximum peak value of the flash at $P_s/P_0=3.2$. At even larger detuning values, the atoms interact less with the light, leading to a decrease of the peak value towards zero. 

When $s_0\leq100$, the transmitted light is strongly absorbed at resonance. It means that the coherently forward-scattered field has the same magnitude as the incident field, but with an opposite phase, leading to the observed values of $P_s/P_0=1$ [see \fig{fig3}(a)].

Above $s_0=100$, the peak values of the flashes start to decrease for all three cases plotted in \fig{fig3}(b). As the transitions of the majority of the atoms  become saturated, the fraction of light that is coherently scattered reduces. This means that $P_s/P_0$ also decreases. This is reflected in the reduction of the peak values of the flashes. 

In \fig{fig3}(b), the experimental points at $\delta/\Gamma=19$ for $s_0\leq100$ are systematically higher than the theoretical prediction.  Some systematic errors in $b_0$ or $\delta$ could  explain these discrepancies, especially when $P_s/P_0$ vary rapidly with respect to either $s_0$ or $\delta/\Gamma$.  This could also explain the discrepancy at $s_0\approx 300$ for the case of $\delta/\Gamma=-2$. 

Finally, we note that our approach neglects backaction of the scattered fields, elastic and inelastic, onto the forward-propagating field due to the nonlinear response of the atomic medium~\cite{binninger2018}. The generally good agreement between the experimental results and the prediction of the model justifies this approximation. The backaction could be important in some cases, such as in parts of a medium not directly illuminated, in an optically thick medium where radiation trapping could happen, or at the weak intensity tail of a Gaussian beam where the saturation parameter could be strongly affected by the scattered light. 

\section{Discussion}\label{sec:discussion}
As discussed in Sec.~\ref{sec:model},
the forward-scattered field is coherent, so it is built up upon elastic events, even in the strong saturation regime where scattering is mainly inelastic. 
We now derive a simple analytical relation linking the forward-scattered power to the elastic scattering power. In general, the existence of such
a
relation is still an open question. However, in the large saturation regime, a simple linear relation between the forward-scattered power $P_s$ and the elastic scattering power $\Pel$ can be found. In this section, we derive this formula and apply it to our experimental data in the large saturation regime, to compare with the numerically calculated value of elastic scattering power.

At first, we remind the reader that by the conservation of energy fluxes in the steady-state regime, the total power scattered by the atomic medium is $\Psca = P_0 - P_t$. $\Psca=\Pel+P_\text{in}$ contains both the elastic $\Pel$ and inelastic $P_\text{in}$ contributions~\cite{cohen2004}. Therefore, the transmitted power is simply related to the total scattered power. 

Now, to find a relation between $P_s$ and $\Pel$, we first consider the resonant case $\delta=0$, and we take the probe beam to be transversally homogeneous within an area of $A$ and an incident saturation parameter of $s_0$. We also consider an atomic medium at zero temperature
with
uniform slab geometry. Finally, we address the problem in
the
weak-absorption limit ($s_t \approx s_0$), which holds either for a medium with $b_0\ll1$ or for a highly saturated medium with large optical depth. By approximating $\ln (s_t/s_0) \approx (s_t-s_0)/s_0$, \eq{wtD} gives 
\begin{equation}
P_t \approx P_0 \left(1-b_0\frac{1}{1+s_0}\right)=P_0-P_{sca}, \label{eq:It1}
\end{equation}
where  $P_0 = AI_{\textrm{sat}} s_0$. Moreover, we know that the elastic contribution reads~\cite{cohen2004}
\begin{equation}
\frac{\Pel}{P_0}=\frac{b_0}{(1+s_0)^2},\label{eq:PelLowOD}
\end{equation}
and the inelastic scattering contribution reads
\begin{equation}
\frac{P_{\textrm{in}}}{P_0}=\frac{b_0s_0}{(1+s_0)^2}.
\end{equation}

From \eq{Et}, we further have the following relation between $P_t$, $P_0$ and $P_s$. 
\begin{equation}
P_t = P_0 + P_s + 2 \sqrt{P_0 P_s}\cos\varphi,
\end{equation}
where $\varphi$ is the phase difference between $E_s$ and $E_0$. In cases where the forward-scattered power is weak, \ie, $P_s\ll (P_0-P_t)$, we can approximate the above equation as: 
\begin{equation}
P_t \approx P_0 + 2\sqrt{P_0P_s}\cos\varphi.\label{eq:It2}
\end{equation}
At $\delta=0$, $\varphi=\pi$ regardless of the value of $s_0$. Thus, we have
\begin{equation}
\frac{P_s} {P_0}\approx \frac{1}{4}\left( \frac{P_0-P_t}{P_0}\right)^2. \label{eq:PsInPt}
\end{equation} 
At resonance, when $s_t\approx s_0$, we have $P_s\ll P_0-P_t$, which justifies our approximation. Using \eqs{It1}{PelLowOD}, we find that the forward-scattered power is proportional to the elastic scattering,
\begin{equation}
\frac{P_s}{P_0} \approx \frac{b_0^2}{4(1+s_0)^2} =\frac{b_0}{4} \frac{\Pel}{P_0}.\label{eq:PsPel}
\end{equation}

We further generalize \eq{PsPel}, including Doppler broadening and small detuning, so the linear relation between $P_s$ and $\Pel$ reads (see details in Appendix~\ref{sec:scattering})
\begin{equation}
\frac{P_s}{P_0} \approx \frac{b_0}{4\cos^2\varphi}\frac{1}{1+4(k\vbar/\Gamma)^2+4(\delta/\Gamma)^2}\frac{\Pel}{P_0}.\label{eq:PsPelkv}
\end{equation}
This final relation is derived for weak attenuation $s_t\approx s_0$, weak forward scattering $P_s \ll P_0 - P_t$ and strong saturation where
\begin{equation}
s_0 \gg 1 + 8(k\vbar/\Gamma)^2 + 4(\delta/\Gamma)^2.
\end{equation}
The above
assumptions allow
us to rewrite the condition $s_t\approx s_0$ as
\begin{equation}
s_0\gg b_0.
\end{equation}
Since we consider $b_0\gg1$, $P_s \ll P_0-P_t$ can also be rewritten as (see Appendix~\ref{sec:scattering})
\begin{equation}
s_0 \gg b_0 (\delta/\Gamma)^2.\label{eq:Inequalities}
\end{equation}

Within the framework of the model discussed in Sec.~\ref{sec:model}, we can also directly compute the total elastic scattering power. Considering only single scattering events, we sum up the velocity-averaged elastic scattering power over all the atoms at different positions in the medium,
\begin{equation}
\frac{P_\mathrm{el}}{P_0} 
= \int_0^{s_0} \frac{{\rm d}s}{s_0} \int_0^{b_0} {\rm d}\zeta \int_{-\infty}^{\infty} {\rm d}v~\frac{g(v) \left[4(\delta-kv)^2/\Gamma^2+1\right]}{\left[4(\delta-kv)^2/\Gamma^2+1+\eta\right]^2}.\label{eq:Pel}
\end{equation} 
The above expression is valid for any values of $b_0$. Here, the steady-state local saturation parameter at a position parametrized by $(\zeta, s)$ is given by
\begin{equation}
\eta(\zeta, s) = s |{\mathcal E}_t(\zeta,s)|^2.\label{eq:locals}
\end{equation}
${\mathcal E}_t$ is the steady-state value of the coherent forward-propagating field.

\begin{figure}
\includegraphics[width=0.5\textwidth]{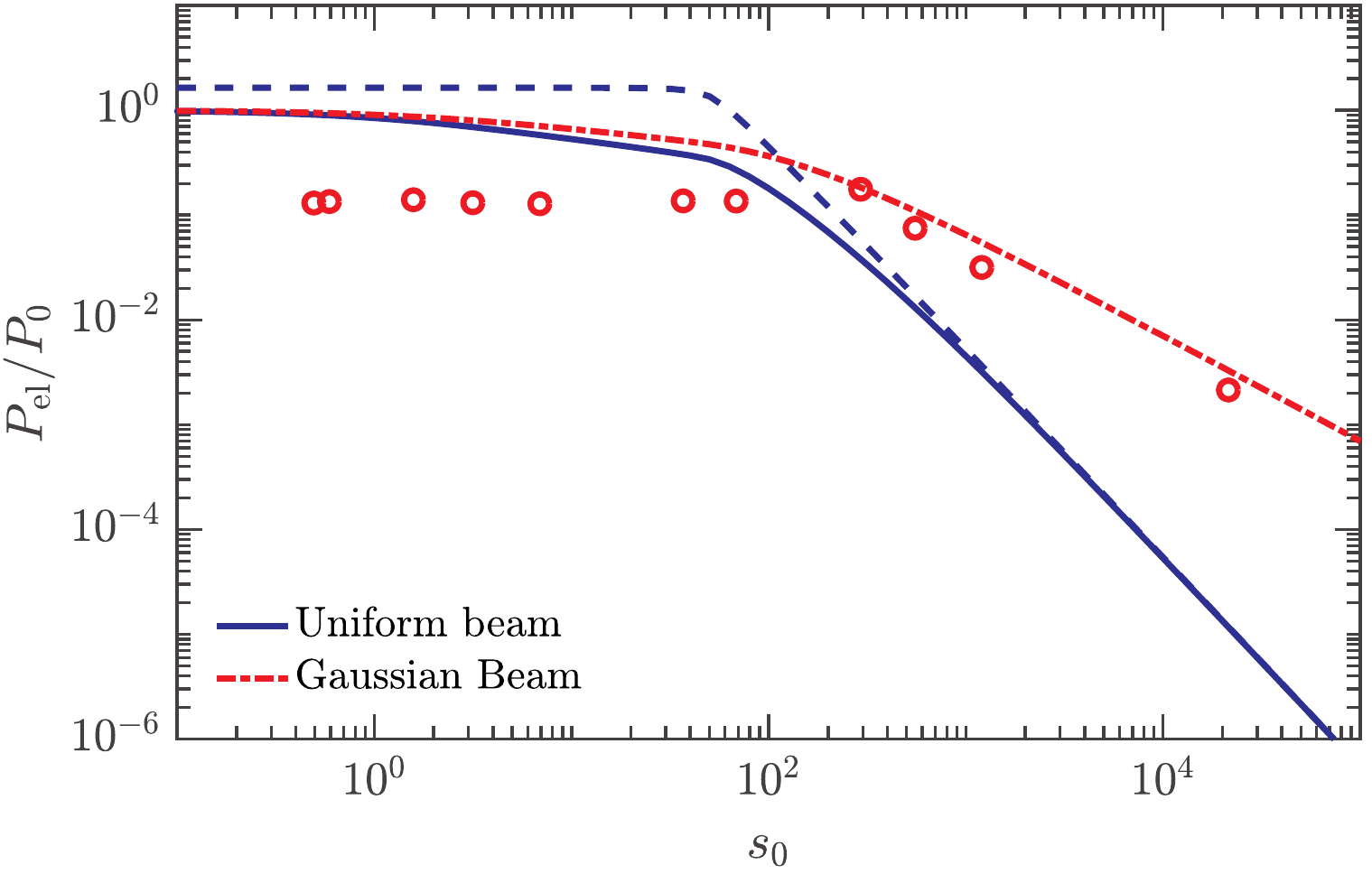}
\caption{Plot showing the
numerically simulated
elastic scattering power
 and the values obtained from 
the experimentally accessible quantity
$P_s$ using \eq{PsPelkv}. The curves are simulated results
for uniform (blue solid) and Gaussian (red dash-dotted) probe beams,
with the details of the simulation provided in the text.
The blue dashed line 
shows
the elastic scattering power computed using the numerically simulated values of $P_s/P_0$ in
\eq{PsPelkv}. The red open circles are the extracted elastic scattering power using the experimental results at $\delta/\Gamma=-2$.
 }\label{fig:fig4}
\end{figure}

We check the validity of \eq{PsPelkv} using \eqs{Js_mod}{Pel} to compute $P_s$ and $\Pel$, respectively. In \fig{fig4}, the 
value of
$\Pel/P_0$,
directly calculated from the numerical simulation according to Eq.~(\ref{eq:Pel}),
is plotted as the solid blue curve, while the numerically calculated $P_s/P_0$  is used in \eq{PsPelkv} to compute the dashed blue curve. Here, $\delta/\Gamma=0$, $b_0=115$, and $k\vbar/\Gamma=3.4$, and thus the two curves converge when $s_0\gg b_0$. For $|\delta|>\Gamma$, the convergence is found for $s_0\gg b_0(\delta/\Gamma)^2$, according to \eq{Inequalities}.

We now aim to extract the elastic scattering power from our transmission measurements. An important difference lies in the Gaussian profile of the probe beam, which requires a transverse integration of the formula given by \eq{PsPelkv}. The factor $\cos^2\varphi$, however, complicates the transverse integration. Fortunately, the value of $\cos\varphi$ saturates to the following at large $s_0$ (see Appendix~\ref{sec:scattering}): 
\begin{equation}
\cos\varphi \approx -\frac{1}{\sqrt{4(\delta/\Gamma)^2+1}}.\label{eq:cosphi}
\end{equation}
This suggests that as long as we are in the large saturation regime, \eq{PsPelkv} can be used as an approximation for a Gaussian probe beam, by taking $\cos\varphi$ to be the value in \eq{cosphi}. Using this approach, we plot the experimental results at $\delta/\Gamma=-2$ (red open circles) in \fig{fig4}. 
For $s_0>b_0(\delta/\Gamma)^2\simeq 460$, a good agreement is achieved between the experimental results and the red dash-dotted curve calculated numerically according to \eq{Pel}. This is despite a value of $P_t/P_0\simeq0.5$ at $s_0\simeq 460$, where the validity of weak absorption approximation can be questioned. In fact, the weak absorption approximation does hold at the center of the Gaussian beam and seems to be sufficient for the agreement. However, the experimental results still underestimate the forward-scattered field. This may be originating from the tails of the Gaussian beam where the saturation parameter is low, and our assumptions do not hold.
 
At large $s_0$, $\Pel/P_0$ for the uniform beam case scales as $s_0^{-2}$, as expected from \eq{PelLowOD}. For the Gaussian beam case, the numerical results of $\Pel/P_0$ scale as  $s_0^{-1}$. The difference between the uniform and Gaussian beams can be understood from the fact that transverse integration is essentially an integration over the saturation parameter [see \eq{Pel}]. Thus, upon transverse integration, the scaling behavior at large $s_0$ changes from $s_0^{-2}$ to $s_0^{-1}$. The experimental estimation of the elastic scattering scales as $s_0^{-0.96(19)}$ for $s_0> 500$, in agreement with the expected behavior.

\section{Conclusion}\label{sec:conclusion}
We have performed experimental studies of the forward-scattered power in 
the
saturated regime. This quantity is extracted from the peak power of the flash, obtained after 
abruptly switching
off the incident laser beam. Our experimental results are well explained by a model based on Maxwell-Bloch equations,  consistent with the fact that only 
coherent elastic scattering
contributes 
to
the forward-scattered light. At large saturation and at the vicinity of the resonance, we derived a relation to compute the total power scattered elastically by the atomic medium
from the measurement of the forward-scattered power. While our relation is valid for a limited range of
parameters,
it can be readily applied to hot or cold atomic
ensembles.
Since the total scattered power is also known from the steady-state transmission
measurement,
this allows us to measure the relative contributions of elastic and inelastic scattering.
This 
bears the
merit of requiring only a simpler transmission measurement in the transient regime,  avoiding more sophisticated measurements of the fluorescence spectrum (see a recent measurement in Ref.~\cite{OrtizGutierrez2019}). 

Finally, several recent works in the linear scattering regime have revealed a density-induced cooperative shift and linewidth broadening of optical transitions~\cite{keaveney2012, jennewein2016,bromley2016, corman2017, jennewein2018, kwong2019}. Recently, the study of collective effects has started to include the saturated regime~\cite{Santo2020}. Our work here could complement such an effort. In particular, in the high-saturated regime, the flash effect could be a simple tool to extract elastic scattering, where strong cooperativity is likely to reinforce it.

\begin{acknowledgments}
The authors thank T. Yang for technical assistance in performing the experiment and data taking, and A. Buchleitner for careful reading of the manuscript. This work was supported by the CQT/MoE funding Grant No. R-710-002-016-271.
\end{acknowledgments}

\appendix

\section{The zero temperature approximation}\label{sec:zeroT}
Under thermal averaging, the scattering cross section of an atom in the saturated regime with a local value of saturation parameter $\eta$ reads
\begin{equation}
\sigma_{\vbar} = \frac{\sigma_0}{1+\eta}\mathrm{Re} \left\{G_{\vbar}(\delta,\eta)\right\},
\end{equation}
with 
\begin{equation}
G_{\vbar}(\delta, \eta) = \sqrt{\frac{\pi}{8}}\frac{\Gamma\sqrt{1+\eta}}{k\vbar} w\left(\frac{\delta+i\Gamma\sqrt{1+\eta}/2}{\sqrt{2}k\vbar}\right),
\end{equation}
and $w(z)$ is the Faddeeva function~\cite{abramowitz1974}. Here, $\mathrm{Im}\{z\}>0$, so the Faddeeva function has the following asymptotic expansion~\cite{gautschi1970}:
\begin{equation}
w(z) \approx \frac{i}{\sqrt{\pi}}\left[\frac{1}{z}+\frac{1}{2z^3}+\frac{3}{4z^5}\right].\label{eq:wasymp}
\end{equation}
For $|\delta + i\Gamma\sqrt{1+\eta}/2|\gg \sqrt{2} k\vbar$, the above asymptotic expansion can be applied throughout the atomic medium. Keeping only the leading term leads us back to the scattering cross section in the absence of Doppler broadening, \ie, 
\begin{equation}
\sigma_{\vbar} \sim \sigma_0\frac{\Gamma^2/4}{\delta^2+\Gamma^2(1+\eta)/4},
\end{equation}
Thus, for sufficiently large $\eta$ or $\delta$ values, the zero temperature limit is  a good approximation. We remind the reader that if $\eta$ is large, then $\eta\approx s_0\approx s_t$.

\section{Derivation of \eq{PsPelkv}}\label{sec:scattering}

For a uniform probe beam, with $s_t \approx s_0$, \eq{Pel} simplifies to
\begin{equation}
\frac{\Pel}{P_0} \approx b_0\int_{-\infty}^{\infty} {\rm d}v~\frac{g(v) \left[4(\delta-kv)^2/\Gamma^2+1\right]}{\left[4(\delta-kv)^2/\Gamma^2+1+s_0\right]^2}.
\end{equation}
After velocity averaging, we find
\begin{widetext}
\begin{equation}
\frac{\Pel}{P_0} \approx b_0\left\{\frac{2+s_0}{2(1+s_0)^2} \mathrm{Re}\left\{ G_{\vbar}(\delta,s_0)\right\} -\frac{s_0}{8(1+s_0)}\left(\frac{\Gamma}{k\vbar}\right)^2\left[1-\mathrm{Re}\left\{G_{\vbar}(\delta,s_0)\right\} \right]+\frac{s_0}{4(1+s_0)^{3/2}}\frac{\Gamma\delta}{(k\vbar)^2}\mathrm{Im}\left\{G_{\vbar}(\delta,s_0)\right\}\right\}
\end{equation}
\end{widetext}
A further simplification is possible by considering $s_0\gg 8(k\vbar/\Gamma)^2$ so that \eq{wasymp} can be applied. Considering further that $s_0\gg1 + 4(\delta/\Gamma)^2$, we get
\begin{equation}
\frac{\Pel}{P_0} \approx \frac{b_0}{s_0^2} \left[1+4\left(\frac{k\vbar}{\Gamma}\right)^2+4\left(\frac{\delta}{\Gamma}\right)^2\right].\label{eq:Pelkvs}
\end{equation}

From \eq{It2}, under the further constrain of weak forward-scattered power $P_s \ll  P_0-P_t$, we find
\begin{equation}
\frac{P_s}{P_0}\approx \frac{(1-P_t/P_0)^2}{4\cos^2\varphi}, 	\label{eq:PsIt}
\end{equation}
whereas the total scattered power is given by
\begin{multline}
\frac{P_0-P_t}{P_0} \approx b_0\int_{-\infty}^{\infty} {\rm d}v~\frac{g(v) }{\left[4(\delta-kv)^2/\Gamma^2+1+s_0\right]^2}\\=\frac{b_0}{1+s_0}\mathrm{Re}\left\{G_{\vbar}(\delta, s_0)\right\}\approx \frac{b_0}{s_0}.
\end{multline}
Upon substituting the above relation into \eq{PsIt}, we get
\begin{equation}
\frac{P_s}{P_0} \approx \frac{b_0^2}{4s_0^2\cos^2\varphi}. 
\end{equation}
Inserting \eq{Pelkvs} in the above equation leads to \eq{PsPelkv}.

We further discuss the range of $\delta/\Gamma$ for \eq{PsIt} to be valid. In this large saturation regime of $s_0 \gg  8(k\vbar/\Gamma)^2$, we can ignore Doppler broadening. We then approximate the effective optical depth $\mathcal{B}$ and the phase shift of the transmitted field $\phi$ as 
\begin{align}
\mathcal{B} &\approx \frac{b_0}{4(\delta/\Gamma)^2+1+s_0},\nonumber\\
\phi &\approx -\frac{b_0 (\delta/\Gamma)}{4(\delta/\Gamma)^2+1+s_0}\label{eq:Bphi}
\end{align}
The forward-scattered field, relative to the incident field at the output surface of the medium, is $E_s/E_0 = \exp(-\mathcal{B}+i\phi)-1$. Thus, the forward-scattered power is given by
\begin{equation}
\frac{P_s}{P_0}= 1 + \exp(-\mathcal{B}) - 2 \exp(-\mathcal{B}/2)\cos\phi.
\end{equation}
The total scattered power is given by 
\begin{equation}
\frac{P_0-P_t}{P_0} = 1 - \exp(-\mathcal{B}). 
\end{equation}
Thus, the inequalty $P_s < P_0 - P_t$ leads to
\begin{equation}
\exp(-\mathcal{B}/2) < \cos\phi.
\end{equation}
If we further have $s_0 \gg 1+4(\delta/\Gamma)^2,\, b_0$, then $\mathcal{B},\, \phi \ll 1$ and we find that the above inequality becomes
\begin{equation}
\phi^2 < \mathcal{B}.
\end{equation}
In our case where $b_0\gg1$, we find the following condition for \eq{PsIt} to be valid:
\begin{equation}
s_0 \gg b_0 \left(\frac{\delta}{\Gamma}\right)^2.
\end{equation}

We further note that in the regime of large saturation that we are considering, the forward-scattered field can approximated as
\begin{equation}
\frac{E_s}{E_0}\approx -\frac{\mathcal{B}}{2} + i \phi,
\end{equation}
from which we find $\cos\varphi$ to saturate to the following value at large $s_0$:
\begin{equation}
\cos\varphi \approx -\frac{1}{\sqrt{4(\delta/\Gamma)^2+1}}.
\end{equation}

\bibliography{biblio}
\end{document}